\pdfoutput=1
\documentclass{article}
\usepackage{arxiv}

\usepackage[utf8]{inputenc} 
\usepackage[T1]{fontenc}    
\usepackage{hyperref}       
\usepackage{url}            
\usepackage{booktabs}       
\usepackage{amsfonts}       
\usepackage{nicefrac}       
\usepackage{microtype}      
\usepackage{lipsum}		
\usepackage{graphicx}
\usepackage{doi}
\usepackage{amsmath,amsfonts}
\usepackage{algorithmic}
\usepackage{algorithm}
\usepackage{array}
\usepackage[caption=false,font=normalsize,labelfont=sf,textfont=sf]{subfig}
\usepackage{textcomp}
\usepackage{stfloats}
\usepackage{url}
\usepackage{verbatim}
\usepackage{graphicx}
\usepackage{cite}
\usepackage{url}
\usepackage{tikz}
\usepackage{xcolor}
\usepackage{pifont}
\usepackage{verbatim}
\usepackage{amsmath}
\usepackage{subfig}
\usepackage{siunitx}
\usepackage{blindtext}
\usepackage{makecell}
\usepackage{multirow}
\usepackage{xcolor}
\usepackage{graphicx}
\usepackage{url}
\usepackage{threeparttable}
\usepackage{fancyhdr}
\usepackage{physics}
\usepackage{nameref}
\usepackage{bookmark}
\usetikzlibrary{chains,scopes,positioning,backgrounds,shapes,fit,shadows,calc,arrows.meta,shapes.geometric}
\usetikzlibrary{quotes,angles}
\usetikzlibrary{calc}
\usetikzlibrary{decorations.pathreplacing,math}
\RequirePackage{amsmath,mathrsfs,amsfonts}
\usepackage{circuitikz}
\definecolor{myNewColorA}{RGB}{29,104,174}
\definecolor{myNewColorB}{RGB}{56,133,196}
\definecolor{myNewColorC}{RGB}{98,164,208}
\definecolor{myRed}{RGB}{216,56,58}
\definecolor{myGreen}{RGB}{84,179,69}
\definecolor{myGrey}{RGB}{145,147,147}
\definecolor{waterColor}{RGB}{38, 82, 140}
\definecolor{colaColor}{RGB}{60, 146, 166}
\definecolor{alcoholColor}{RGB}{242, 211, 153}
\definecolor{lemonColor}{RGB}{217, 171, 115}
\definecolor{oilColor}{RGB}{166, 118, 101}
\definecolor{pepsiColor}{RGB}{237, 178, 32}
\definecolor{spriteColor}{RGB}{183, 39, 62}
\definecolor{solutionColor}{RGB}{0, 114, 189}
\colorlet{Ecol}{orange!90!black}
\colorlet{EcolFL}{orange!80!black}
\colorlet{veccol}{green!45!black}
\colorlet{EFcol}{red!60!black}
\colorlet{pluscol}{red!60!black}
\colorlet{minuscol}{myNewColorA}
\usepackage{bm}
\def\systemname{\emph{RainfalLTE}}
\def\netname{\emph{RainNet}}

\hypersetup{
colorlinks=true,
linkcolor=black,
citecolor=waterColor
}

\title{RainfalLTE: A Zero-effect Rainfall Sensing System\\ Utilizing Existing LTE Infrastructure}

\author{Pengfei Shi
\And Fei Shang
\And Haohua Du
}

\renewcommand{\shorttitle}{RainfalLTE}

\hypersetup{
pdftitle={RainfalLTE: A Zero-effect Rainfall Sensing System\\ Utilizing Existing LTE Infrastructure},
pdfauthor={Xianbin Jiang, Fei Shang, Haohua Du, Panlong Yang, Xing Guo, Lihong Liang, Yuanting Zhang, Xiang-Yang Li},
pdfkeywords={IoT, wireless sensing, rainfall sensing, LTE
},
}

\begin{document}
\maketitle

\begin{abstract}
	Environmental sensing is an important research topic in the integrated sensing and communication (ISAC) system.
  Current works often focus on static environments, such as buildings and terrains.
  However, dynamic factors like rainfall can cause serious interference to wireless signals.
  In this paper, we propose a system called \systemname\ that utilizes the downlink signal of LTE base stations for device-independent rain sensing.
  In particular, it is fully compatible with current communication modes and does not require any additional hardware.
  We evaluate it with LTE data and rainfall information provided by a weather radar in Badaling Town, Beijing
  The results show that for 10 classes of rainfall, \systemname\ achieves over 97\% identification accuracy.
  Our case study shows that the assistance of rainfall information can bring more than 40\% energy saving, which provides new opportunities for the design and optimization of ISAC systems.
\end{abstract}

\keywords{IoT, wireless sensing, rainfall sensing, LTE}

\section{Introduction}
Describing the impact of the environment on wireless signals is crucial for communication and sensing systems.
 Traditionally, we often use random modeling based on certain parameters to describe wireless channels such as Rayleigh fading and Rician fading~\cite{tseFundamentalsWirelessCommunication2011}.
 Despite their simplicity and ease of handling, these models are limited in their ability to optimize communication and sensing systems for specific environments due to their reliance on statistical information alone, lacking concrete details about the environment.

 Many excellent works~\cite{zengEnvironmentAware6GCommunications2020} have attempted to generate channel knowledge maps (CKM) based on geographical information including building location, altitude, and terrain.
 They integrate this with traditional communication system design schemes to provide a larger space for system optimization.
 However, they only focus on static geographic information (e.g. the location and height of buildings), which will not change significantly for a long time after one measurement and calibration.
 unfortunately, \textbf{the performance of communication and sensing systems still suffers from severe interference by dynamic environments}, such as a reduction in the coverage area of 5G base stations by about 40\% even with light rain.

 Traditional rainfall measurement means include rain gauges and weather radar, but \textbf{integrating them with existing wireless systems is costly}.
 Currently, there are many excellent works~\cite{gossetImprovingRainfallMeasurement2016,overeemRetrievalAlgorithmRainfall2016} studying the feasibility of using LTE base stations for rainfall sensing.
 However, they are all based on air-to-air links between base stations, a communication method that is gradually being replaced by underground optical cables.
 This makes them \textbf{incompatible with the current communication systems}.
 Beritelli et al.~\cite{beritelliRainfallEstimationBased2018}
implement rainfall classification based on the downlink between the base station and the terminal, but they are implemented based on a single base station so that \textbf{they depend on specific equipment features and configurations}.

 \begin{figure}[t]
    \centering
    \includegraphics[width=0.7\linewidth]{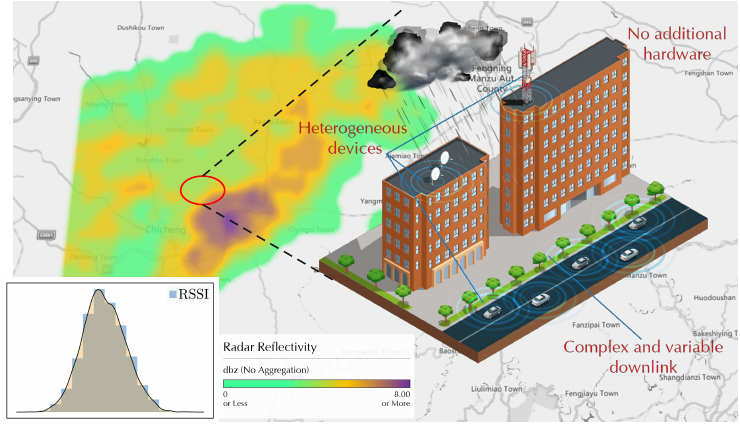}
    \caption{Sensing the rainfall based on LTE downlink signals.}
    \label{fig:intro1}
    \vspace{-1em}
\end{figure}

 In order to be used in a real system, it needs to meet the following three requirements:
 \begin{itemize}
    \item \textbf{Zero hardware expansion.} The system should be able to sense rainfall without any additional hardware, so as to avoid high equipment and deployment costs for the system.
    \item \textbf{Zero effect.} The sensing system must be compatible with the current base station communication mode, and it cannot interfere with the normal operation of the communication system.
    \item \textbf{Device-independent.} Because of the huge variety in models and types of our base stations and terminals, we need to build sensing systems that do not depend on specific device characteristics.
 \end{itemize}

 In this paper, we propose \systemname, a system that uses the communication link between base station and terminal devices to sense rainfall.
 In particular, \systemname\ is independent of the specific base station configurations, which makes \textbf{it promising to provide beneficial information for the optimization of large-scale ISAC systems}.
 Rainfall sensing based on LTE downlink signals is motivated by a fundamental observation: rain causes attenuation in the LTE signal.
However, there are three challenges that need to be addressed:

(1) \textbf{The available information is scarce}. Unlike meteorological radar or rain gauges that can provide high-precision (sub-millimeter level) rainfall information, the data reported by base stations only contain records about the received signal strength. Not only is there a lack of information on the signal phase, but more seriously, the received intensity values are discrete integer data. 

(2) \textbf{The interference of the environment is complex and variable.}
Compared to the air-to-air link, the communication link between the base station and the terminal equipment is severely affected by environmental interference.
To make matters worse, in addition to the moisture in the raindrops that will affect the attenuation of the signal, the antenna wetness caused by dew~\footnote{Previous studies~\cite{uptonMicrowaveLinksFuture2005} have shown that the received signal strength fluctuation caused by antenna wetness and so on is as much as \SI{0.5}{dB}.} and so on will also cause interference to the rainfall sensing. However, these interferences are very strong randomness.

(3) \textbf{Lack of prior knowledge about the device}. Due to the interference of the base station's downlink signal by the device status, including transmission and reception power, distance between transmitter and receiver, antenna orientation, type of device, etc. However, this information is unknown.

\textbf{Solutions.}
(1) We extract statistical characteristics such as probability distribution for rainfall sensing through stochastic signal modeling. According to the law of large numbers, for time-invariant random signals, when the sample size is sufficiently large, its statistical characteristics tend to stabilize, which suppresses the impact of device differences on the sensing results.
(2) Noting that the weather conditions within a few kilometers are similar, we form a graph with several adjacent base stations and then use graph neural networks to extract their overall features. This helps to suppress interference caused by certain base stations due to localized dampness, among other reasons.
(3) Unlike meteorological radar and satellite-borne radar, the downlink of LTE signals is closely related to pedestrian activities. Based on this, we attempt to extract information related to human activities such as indoor and outdoor device information from LTE data to enhance sensing effects.

\textbf{Contributions:} The main contributions in \systemname\ summarize as follows:

(1) We design \systemname, which can use LTE downlink signals to sensing the rainfall and provide beneficial information for the optimization of a wide range of ISAC systems.

(2) Unlike traditional approaches, we treat signals as random variables and extract features from their probability density functions, which suppresses the interference caused by device differences. Furthermore, we attempt to mine information related to human activities to assist in rainfall sensing, thereby further enhancing the system's sensing upper limit.

(3) We evaluate based on LTE data and gridded radar data in Badaling town, Beijing. We divide the rainfall into 10 levels, with a minimum rainfall of 0 and a maximum rainfall of moderate rain. The results show that the accuracy of sensing rainfall category is over 97\%.
Our case study shows that rainfall information can help ISAC systems solve over 40\% of the energy.

The rest of the paper is organized as follows.
In Section~\ref{sec:pre}, We present the necessity of rainfall sensing.
We introduce the components of our system in Section~\ref{sec:overview}.
We present the basic model of the system and how to extract features and perform accurate rainfall sensing in Section~\ref{sec:system_design}.
Implementation and evaluation are presented in Section~\ref{sec:implementation} and Section~\ref{sec:eva}.
Finally, we discuss the related work in Section~\ref{sec:related_work}.
\section{Why do LTE base stations need to perceive rainfall information?}
\label{sec:pre}

\textbf{The performance of wireless communication systems is influenced by the surrounding environment}, and modeling and analyzing it is crucial for the design, analysis, and optimization of wireless communication.
Noting that base stations are not easily moved after deployment, researchers attempt to integrate the channel knowledge map~\cite{zengEnvironmentAware6GCommunications2020} (CKM) constructed based on geographical environment information (such as buildings) into wireless channel modeling, in order to compensate for the shortcomings of traditional random models~\cite{nurmela2015deliverable} (such as the shadow fading and small-scale fading) that are difficult to optimize at a fine-grained level for the specific environment where the base station is located.
However, \textbf{current CKM only focuses on static geographical information and is difficult to dynamically adjust according to weather} (e.g. rainfall), which makes it difficult for us to optimize the system to suppress the significant impact of weather changes.

In this section, we first analyze the significant impact of rainfall on LTE base stations. Then, we explain how to combine rainfall information with CKM for channel optimization.

\subsection{Rainfall significantly reduces base station coverage.}
\begin{figure*}[t]
\centering
\subfloat[]{
\includegraphics[width=0.28\textwidth]{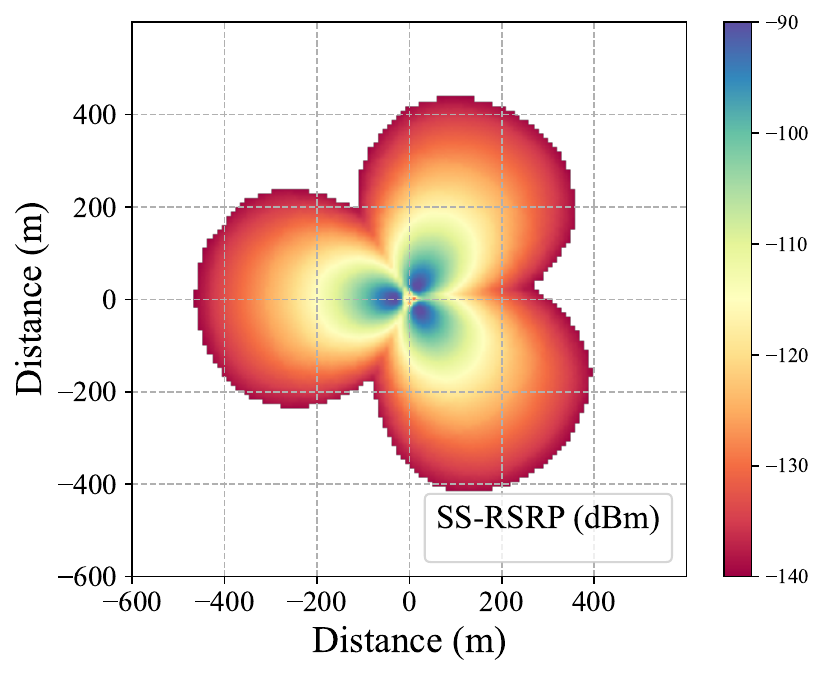}
\label{fig:base_station1}
}
\hfill
\subfloat[]{
\includegraphics[width=0.28\textwidth]{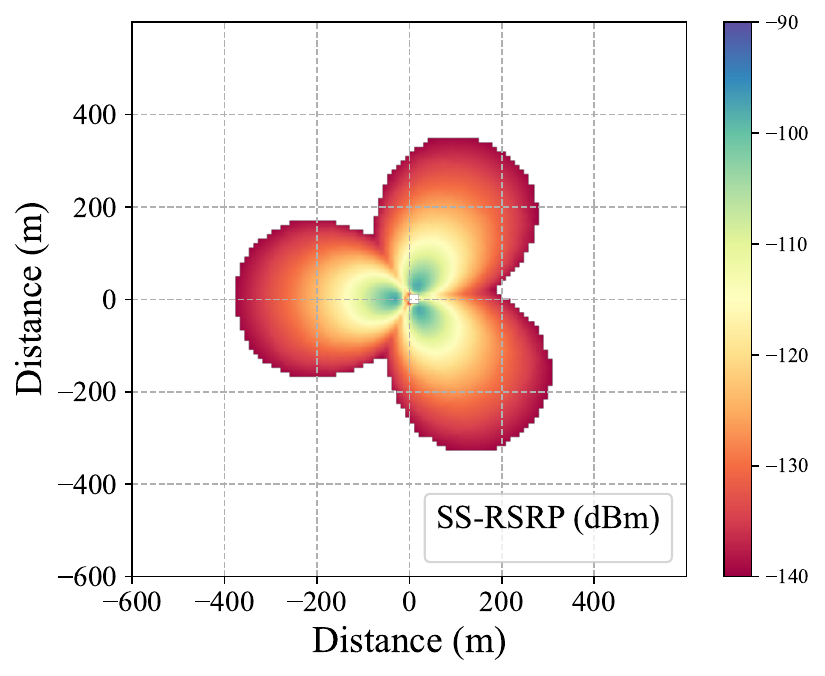}
\label{fig:base_station2}
}
\hfill
\subfloat[]{
\includegraphics[width=0.34\textwidth]{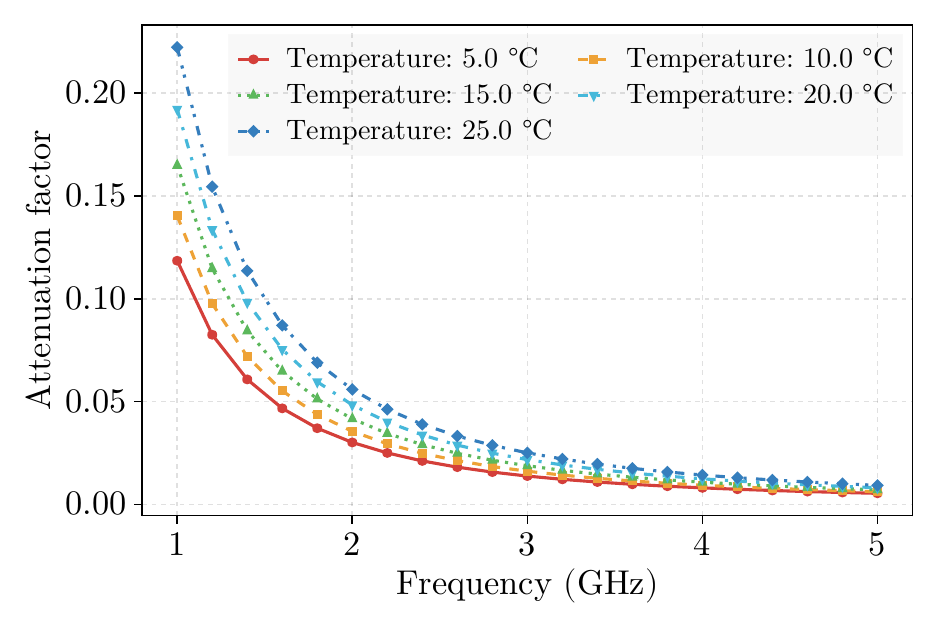}
\label{fig:base_station3}
}
\caption{Rainfall causes significant interference to LTE signals. (a) The coverage area of the base station signal on a sunny day. (b) Rainy weather causes a reduction in the coverage area by about 40\%. (c) Signals in different frequency bands are sensitive to rain fading to varying degrees.}
\label{fig:base_station}
\end{figure*}

\textbf{Rainfall will exacerbate the path attenuation of LTE signals.}
 Our analysis based on real data shows that even light rain will cause a coverage area reduction of about 40\% for base stations.
 Moreover, since the attenuation factor of  is frequency-dependent, LTE signals in different frequency bands have different sensitivity to rain attenuation.
 These results indicate that it is valuable to utilize base station LTE signals to sense rainfall for wireless channel optimization.

 We first simulated the impact of rainfall on the signal coverage area of base stations based on actual data.
 The coverage area of a 5G base station is mainly affected by antenna gain, antenna radiation pattern, transmit power, and path loss.
 Among them, we believe that only the path loss is interfered with by rain.
 Not generally, we consider an array composed of three directional antennas whose radiation patterns divide the horizontal plane into three equal parts.
 We use open-source measured data in the 5G n77 band~\cite{1:misc} to fit the path loss function.
 Figure~\ref{fig:base_station1} shows the coverage area of the base station signal without rain fading.
 Previous research work~\cite{beritelliNeuralNetworkPattern2017} shows that for 4G LTE signals, \textbf{the rain-induced attenuation is about \SI{8.45}{dBm}, which causes a reduction of about 40\% in the coverage area of the base station} as shown in Figure~\ref{fig:base_station2}.
 This result indicates that rainfall has a serious potential interference on the performance of communication systems.

 In addition, previous research works~\cite{PackquIDInpacketLiquid,shangContactlessFineGrainedLiquid2024} have shown that the attenuation of electromagnetic signals in liquids is frequency-dependent, which provides more opportunities for rain sensing at base stations and optimization of communication systems.
 Based on empirical formulas\cite{kaatzeComplexPermittivityWater1989}, we simulated the attenuation factors\footnote{The width of the material needed to decay the strength of the electromagnetic field to $1/e$ of its original value~\cite{dhekne2018liquid}} of water at different frequencies and temperatures, as shown in Figure~\ref{fig:base_station3}.
 The results indicate that at a frequency of \SI{5}{GHz}, the signal's attenuation rate in water is more than 10 times faster than at \SI{1}{GHz}.

\subsection{Typical usage scenarios}
Rainfall information can be used for a variety of purposes, such as cell site selection, adaptive communication, etc. 
It is particularly attractive in the following scenarios.

(1) \textbf{System performance optimization.} 
It is expected that 6G will use large-scale antenna arrays and wide bandwidth to provide communication and sensing mediums for numerous devices, which will lead to an increase in computational complexity for pilot-based channel estimation.
On the other hand, rainfall information can be used as prior knowledge to reduce the difficulty of channel estimation, and even combined with CKM to achieve pilot-free channel response estimation.
Moreover, the sensing of environmental information such as rainfall can facilitate the implementation of dynamic configurations at base stations, achieving effects such as finer-grained beamforming and energy conservation.

(2) \textbf{Improved terminal sensing capability.}
Ubiquitous sensing at the terminal side is playing an increasingly important role.
In many scenarios, real-time acquisition of accurate rainfall information can help further fine-tune system performance.
For instance, in autonomous driving scenarios, rainfall information can be used to infer visibility, road friction coefficient, and other information to optimize the performance of driving algorithms.
However, current sensing terminals often rely on third-party APIs to obtain regional rainfall information, which are often coarse in terms of space and accuracy: spatial resolution is often dozens of kilometers squared, and accuracy is often based on weather classifications such as "light rain" or "moderate rain".
This makes it difficult to provide effective information for ubiquitous infinite sensing systems. Moreover, since the coverage area of base stations is often less than \SI{3}{km} (for 4G signals) or even just a few hundred meters (for 5G signals), if we could provide fine-grained rainfall sensing at the base station, it would offer more beneficial information for the terminal's sensing tasks.
\section{System overview}
\label{sec:overview}

\begin{figure}
    \centering
    \includegraphics[width=0.6\linewidth]{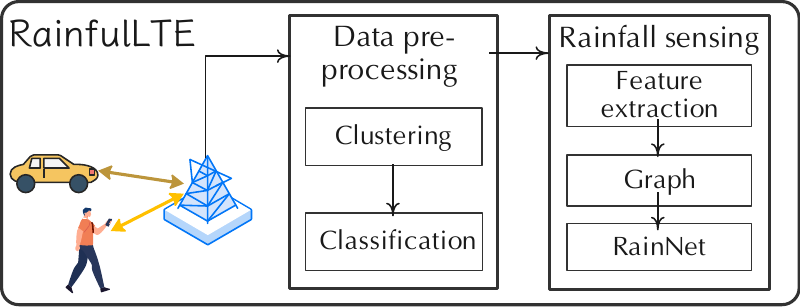}
    \caption{Overview of \systemname.}
    \label{fig:overview}
\end{figure}

\begin{figure*}
    \centering
    \subfloat[]{\includegraphics[width=0.4\linewidth]{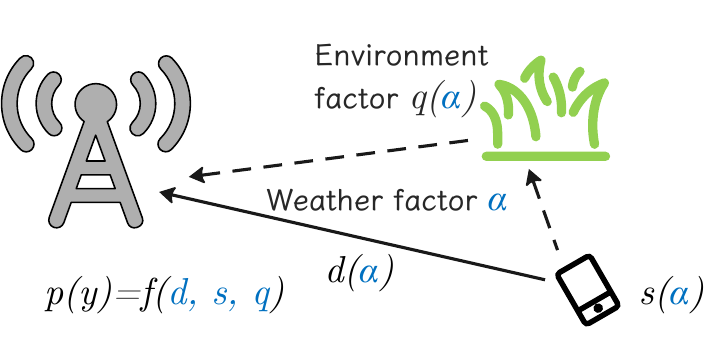}\label{fig:model}}
    \hfill
    \subfloat[]{\includegraphics[width=0.28\linewidth]{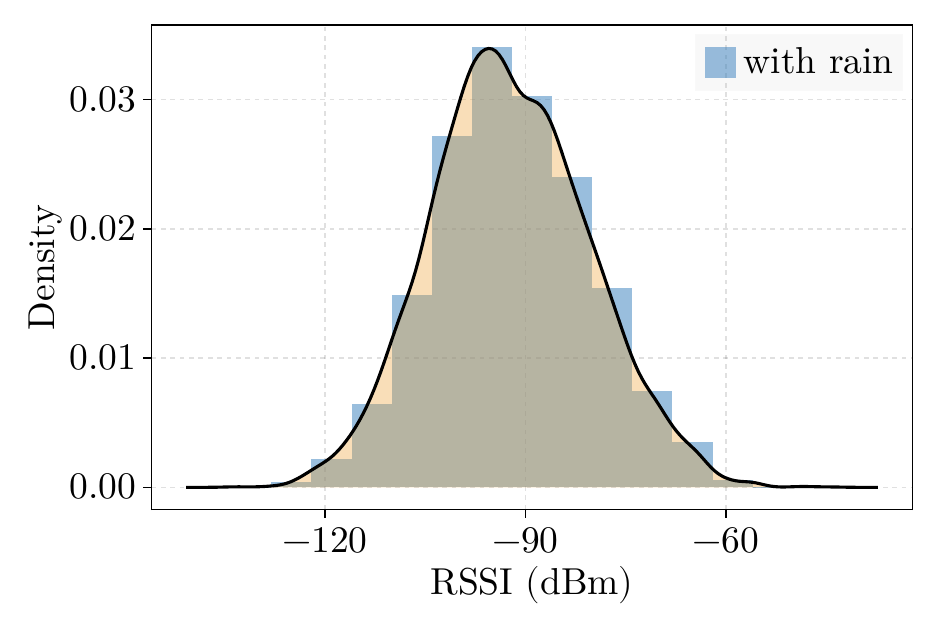}\label{fig:model_rain}}
    \hfill
    \subfloat[]{\includegraphics[width=0.28\linewidth]{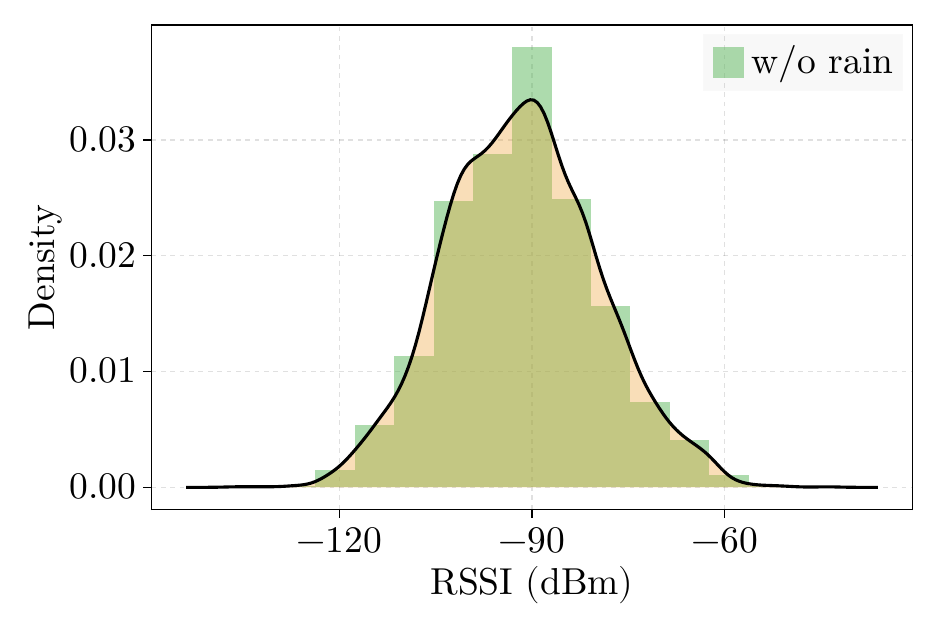}\label{fig:model_no_rain}}
    \caption{Basic model of the system. (a) The probability density distribution of received signal strength is affected by the environment, distance, and device status, all of which are related to weather. (b) The RSSI distribution during rainfall. (c) The RSSI distribution when it does not rain.}
\end{figure*}
Utilizing LTE signals to perceive rainfall, an intuitive approach is to use historical data to fit the parameters of the rain attenuation formula~\cite{RECOMMENDATIONITUR8383}.
However, due to factors such as multipath interference and significant differences among different terminals (for instance, communication differences existing in mobile phones of different types and orientations), the fitting accuracy is not satisfactory.
Recently, some studies~\cite{shangLimitsSensingCapability2024} on sensing channels have shown that even with excellent data preprocessing schemes, it is difficult to greatly improve accuracy when feature correlation is poor.
So we don't use the pattern of data preprocessing and then processing with physical formulas~\cite{RECOMMENDATIONITUR8383}.
Instead, \textbf{we construct features using statistical properties and use graph neural networks for rainfall sensing}.
The fundamental idea originates from a fact: rainfall can cause changes in factors such as rain attenuation and terminal states (for example, more mobile phones are moved indoors after it rains), which leads to changes in the probability distribution of base station LTE signals.
For the same base station, these changes are relatively stable.
In addition, we note that within a region of several kilometers in diameter, the weather conditions are essentially the same.
Therefore, we form a graph consisting of several adjacent base stations, and then use graph neural networks to extract their overall features for rainfall sensing.
\systemname\ is mainly composed of the following three parts:
\begin{itemize}
    \item \textbf{Data preprocessing}: Due to the lack of base station location information, we first cluster LTE signals according to longitude and latitude to approximate the position of equivalent base stations by clustering centers. For the data of LTE downlink, we classify it according to signal categories (including 4G and 5G SA) and remove outliers for subsequent processing. For radar data, we interpolate it to obtain weather information at the base station.
    
\item \textbf{Rainfall sensing}: We treat a base station as a node, extracting statistical characteristics and human flow information from the signal as node features to suppress interference caused by device and environmental differences. Adjacent nodes are connected to form a complete graph, which helps us utilize spatial correlation to suppress interference caused by factors such as antenna dampness. Then, we use \netname\ for rainfall result sensing.
\end{itemize}
\section{System Design}
\label{sec:system_design}

\subsection{Basic model.}
\label{sec:basic_model}

We do not directly build a model of rainfall and signal strength, which suffers from serious interference of environmental and equipment differences.
Instead, \textbf{we focus on the probability density distribution of signal strength over time}, based on a face that the distribution of environment and equipment remains relatively stable in a short period of time.
For example, in about half an hour, if there is no change in weather, in a square with many users, although different users have different devices and they are constantly flowing, they tend to be convergent from the distribution point of view\cite{horanontWeatherEffectsPatterns2013}.
Based on the law of large numbers, when the number of samples is large enough, we can estimate the probability density of the sample with small error.
\textbf{This helps us suppress the influence of equipment differences by extracting features from the probability distribution}.

We consider a model as shown in Figure~\ref{fig:model}.
At a distance d from the base station, there is a terminal that communicates with the base station. The received signal strength is mainly affected by the state of the terminal itself (such as orientation, transmission power, etc.), the surrounding environment (such as multipath, etc.), transmission distance, and weather (rainfall can change path attenuation and the refractive index of the surrounding environment).
Therefore, for the RSSI $Y\sim p(y)$ of the base station, we have:
\begin{equation}
    p(y) = f(d,s,q) + N,
    \label{eq:weather}
\end{equation}
where $d$ is the distance between the terminal and the base station, $s$ is the state of the terminal, $q$ is the environment factor, $N$ is the noise of the system.
\textbf{These factors change significantly when rainfall regimes change.}
Specifically, after rainfall, more pedestrians will move from outdoors to indoors, which will change the probability distribution of distance $d$; raindrops will alter the transmission loss of electromagnetic signals in the air and the refractive index of the ground, changing environmental factors $q$; the state $s$ of the terminal will also change after rainfall, for example, the distribution of vehicle driving speed changes, introducing different Doppler shifts to $s$.
Therefore, we rewrite Equ.~\eqref{eq:weather} as
\begin{equation}
    p(y) = f(d(\alpha),s(\alpha),q(\alpha)) + N,
    \label{eq:weather2} 
\end{equation}
where $\alpha$ is the weather factor.

In a short period of time (e.g. half an hour), if the weather doesn't change, we believe that $d$, $s$, $q$, $N$, are all stationary random variables, i.e. $Y$ is a stationary stochastic process.
When the weather changes, the probability distribution of $Y$ will also change.
Therefore, we can design features based on the probability distribution $p(y)$.

\textbf{This approach has two advantages}: (1) Since we are more concerned with the probability distribution rather than specific values, there is no need to consider the differences in transmission power among different base stations; (2) Since the random variable is stationary, using its statistical characteristics is more conducive to extracting features that reflect the properties of the random variable.

\begin{figure*}[t]
    \centering
    \subfloat[]{\includegraphics[width=0.28\linewidth]{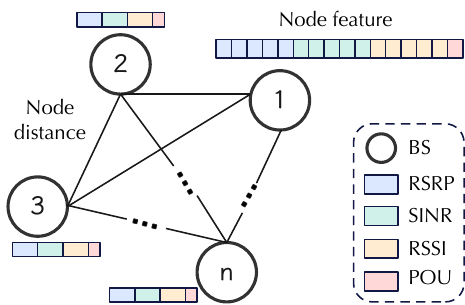}\label{fig:graph}}
    \hfill
    \subfloat[]{\includegraphics[width=0.65\linewidth]{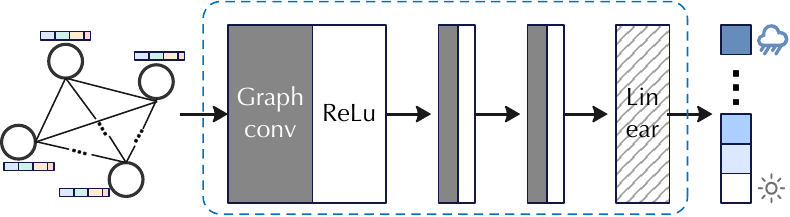}
    \label{fig:graph_conv}}
    \caption{Rainfall sensing network. (a) We construct a graph with adjacent base stations, using signal strength information and user behavior data to form node features. (b) We utilize \netname\ to extract common features of the nodes for fine-grained rainfall sensing.}
\end{figure*}

\subsection{Practical issues.}
Although the model in Sec.~\ref{sec:basic_model} has many advantages, there is a practical problem to be faced in actual use: in addition to rainfall, \textbf{some ``local factors" still have a significant impact on weather factor} $\alpha$.
Previous studies~\cite{overeemRetrievalAlgorithmRainfall2016,uptonMicrowaveLinksFuture2005} have shown that the fluctuation of RSSI caused by factors such as antenna wetness is about \SI{1}{dB} in non-raining conditions, which makes the accuracy and stability of rainfall prediction results seriously damaged when using air-to-air base station communication signals~\cite{overeemRetrievalAlgorithmRainfall2016}.
Moreover, unlike in air-to-air communication where the distance and geographical environment between the transmitter and receiver do not significantly change, \textbf{the interference experienced in communication between the base station and the terminal is more variable}. For instance, the posture of the terminal (such as orientation and azimuth) is more diverse, and there is interference from Doppler frequency shifts (movement of vehicles), etc.

\subsection{Rainfall sensing network.}

To perform accurate rainfall observations without interference from ``local factors" (including the wetness of antennas, etc.), an intuitive idea is that rainfall is similar in a small area (a few square kilometers)~\cite{hennAssessmentDifferencesGridded2018}.
Therefore, we form a graph by combining several adjacent base stations, and then \textbf{use the graph convolutional network to extract their common features for rainfall sensing}.

\textbf{Base station location estimation.}
An intuitive idea is to use the location information of base stations to filter out several base stations closest to it and form a graph for subsequent training.
Unfortunately, the actual data obtained at the base station does not include the position information of the base station itself.
Luckily, terminal devices will report their own longitude and latitude information.
Given that the transmission distance of LTE signals is limited (the transmission distance of 4G signals is about 1-\SI{3}{km}, and the transmission distance of 5G signals is usually less than \SI{300}{m}, this distance may be further reduced due to factors such as building obstructions), we believe that the distance between the terminal device and the base station it communicates with will not be too far.
Therefore, we perform nearest neighbor clustering on LTE dataset $L$ according to latitude and longitude, with the distance calculated using the spherical distance formula, resulting in $m$ different clusters that is $l_1, l_2,\ldots,l_m$.
We use these $m$ cluster centers $b_1, b_2,\ldots,b_m$ as the approximate locations of the base stations.

\textbf{Extract node features.}
We treat a base station as a node, and then extract the features contained in the node from the data.
For each LTE data, we focus on the following categories of data:
\begin{itemize}
\item \textit{RSRP}: Downlink signal strength measurement.
\item \textit{SINR}: Downlink Signal Interference-to-Noise Ratio.
\item \textit{RSSI}: Received signal strength.
\item \textit{POU}: Proportion of outdoor users.
\end{itemize}
In this context, RSRP, SINR, and RSSI are quantities related to signal strength, the distribution of which is influenced by rainfall according to the model introduced in Sec.~\ref{sec:basic_model}.
Furthermore, \textbf{human activities are affected by weather conditions\cite{horanontWeatherEffectsPatterns2013}, and the number of outdoor users has a certain correlation with rainfall.}
Therefore, we also consider the proportion of outdoor users as a feature.

(1) \textbf{Obtain the probability distribution of RSRP, SINR, and RSSI.}
Noting that the values of RSRP, SINR, and RSSI in the base station's messages are all integer types, we estimate them using the histogram method.
We take RSRP data as an example for explanation, and the processing methods for SINR and RSSI data are similar.
Specifically, we first obtain the maximum and minimum values of RSRP data in LTE dataset $L$, and we divide the whole numerical interval into $k$ sub-intervals.
For the $i$-th node, we believe that the base station is located at $b_i$ and its LTE data belong to set $l_i$.
For the RSRP data in $l_i$, we calculate the frequency in each interval.
During this time, we have $m_i$ measurements of the RSRP, which can be represented as a sequence of random variables $\hat{RSRP}$.
According to the law of Uncle Bernoulli, we have
\begin{equation}
    \lim_{m_i\to\infty} p(\hat{RSRP}) \to p(RSRP),
\end{equation}
where $\hat{RSRP}$ is the sequence of measurements of the RSRP.
Then this $k$-dimensional vector is used as the probability distribution of the RSRP data.

\begin{figure*}
    \centering
    \subfloat[]{\includegraphics[width=0.31\linewidth]{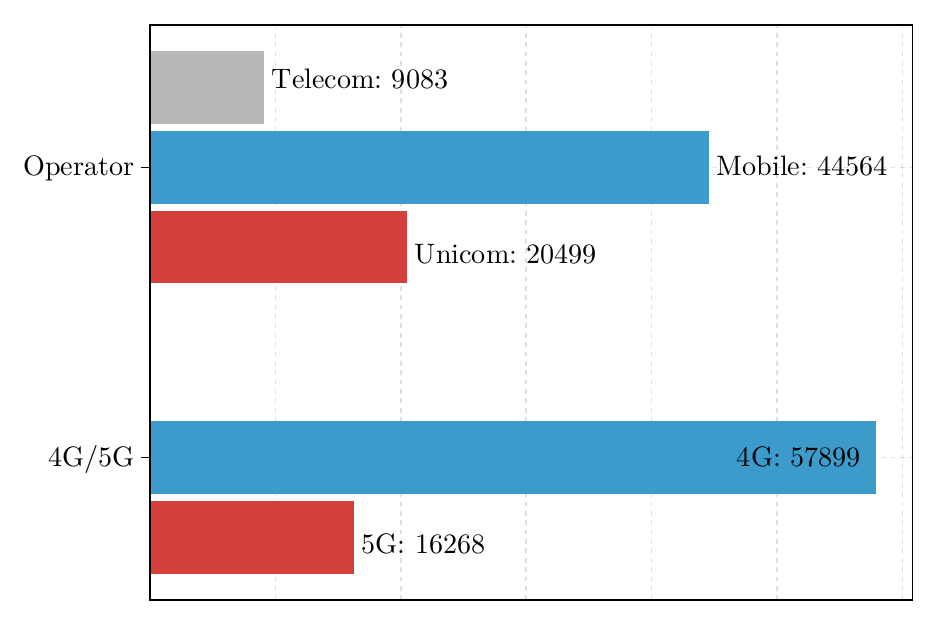}\label{fig:implementation1}}
    \hfill
    \subfloat[]{\includegraphics[width=0.31\linewidth]{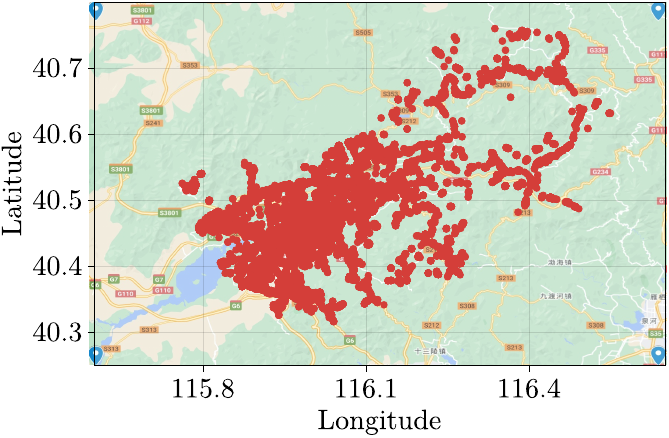}\label{fig:implementation2}}
    \hfill
    \subfloat[]{\includegraphics[width=0.31\linewidth]{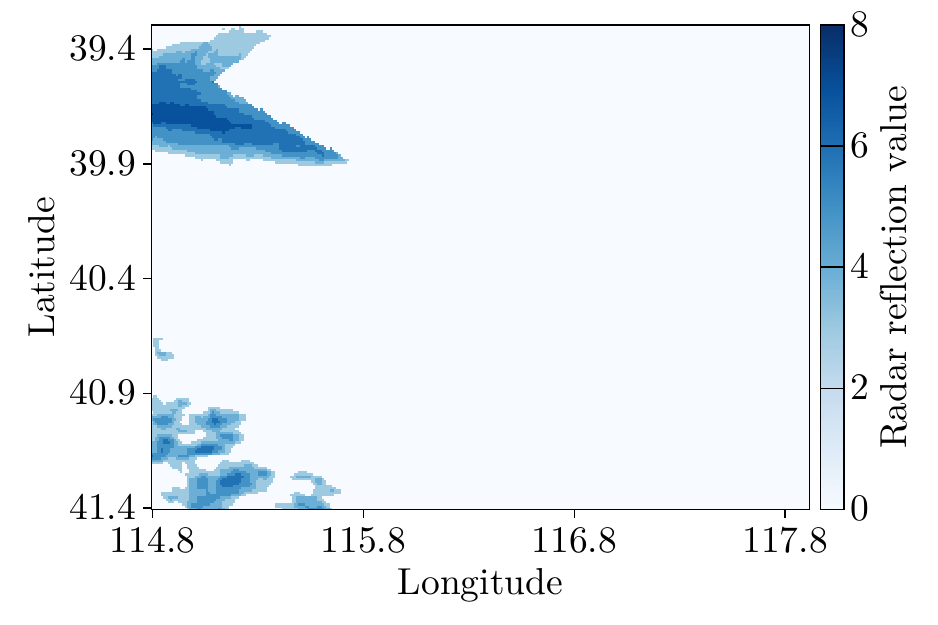}\label{fig:implementation3}}
    \caption{Implementation. (a) Classification of LTE data. (b) Spatial distribution of LTE data. (c) Schematic diagram of meteorological radar data.}
    \label{fig:implementation}
\end{figure*}
(2) \textbf{Obtain the probability distribution of POU.}
Since the LTE messages contain information about whether the terminal device is indoors or outdoors, we directly use the proportion of outdoor users as a feature.

Therefore, the dimension of node features is $3k+1$, which includes three $k$-dimensional vectors to represent the probability distribution of RSRP, SINR, and RSSI respectively, and one data to characterize POU.

\textbf{Form a graph.}
As shown in Figure.~\ref{fig:graph}, we form a complete graph using $n$ adjacent base stations.
For a base station located at $m_i$, we form a graph with it and the nearest $n-1$ base stations.
Each base station serves as a node in the graph.
The characteristics of the nodes are constructed as described above, which include information on RSRP, SINR, RSSI, and POU.
We connect each pair of nodes to form an undirected complete graph, with the edge characteristics being the distance between nodes.

\textbf{Graph convolutional network.}
The structure of \netname\ is shown in Figure~\ref{fig:graph_conv}.
For a complete graph, we first pass it through three layers of convolutional units, each consisting of one layer of graph convolution with ReLU as the activation function.
After that, we add a linear layer which outputs a vector to represent the likelihood of different rainfall classes.
The labels of the graph are derived from grid radar data, and we divide the radar data into $r$ categories at equal intervals from the minimum to maximum values.
The loss function is given by:
\begin{equation}
    \mathcal{L}=\mathcal{L}_{cro}(\hat{y},y),
    \label{eq:loss}
\end{equation}
where $\mathcal{L}_{cro}$ is the cross-entropy loss function, $\hat{y}$ is the output of the network, and $y$ is the label.
\section{Implementation}
\label{sec:implementation}

\subsection{Dataset.}
\label{sec:dataset}
The data we used consists of two parts: the data from LTE base stations, which is used to construct the graph; and the data from weather radars, which is used to construct the labels.
The LTE data is sourced from Yanqing District, Beijing, with the date dimension being October 3, 2022.
It contains a total of 74,167 data entries, including 57,899 4G data entries and 16,268 5G SA data entries.
These data are include four operators, namely China Mobile, China Unicom, China Telecom, and China Broadnet.
These data are supplied by a company.
The category information of these data is shown in Figure.~\ref{fig:implementation1}.\footnote{Since there are only 21 data is from China Broadnet, we don't label them in the figure}
Figure~\ref{fig:implementation2} shows the latitude and longitude location of each data.
Meteorological data is provided by a weather company. 
They monitor the weather using meteorological radar, inferring the amount of rainfall through radar echoes.
The data is recorded every half hour. The radar echo data from 00:00 to 00:30 on October 3, 2022, is shown in Figure~\ref{fig:implementation3}.

\subsection{Details of data processing.}
We cluster the LTE data based on their longitude and latitude information into 100 categories, which we consider as there being 100 base stations in the region.
Then, we form a graph with adjacent 9 base stations, i.e., there are 9 nodes in one graph.
For RSRP, SINR, RSSI, we characterize their probability distributions respectively with a 5-dimensional vector, making the node features $3\times 5+1=16$ dimensions.
For the graph convolutional network, the input and output channel numbers of the first layer of the convolutional network are respectively 9 and 64, while the remaining two layers have a channel number of 64.
For radar data, we equally divide it into 10 intervals, making the category number dimension 10.

\subsection{Software.}
We implement the data processing with Python and Julia scripts.
Moreover, we build \netname\ with PyTorch Geometric\cite{feyFastGraphRepresentation2019} that is one of the most popular frameworks for graph neural networks.
The training process is conducted on a server with an Intel(R) Xeon(R) Silver 4210R CPU, with Ubuntu 22.04 as the operating system.
The overall training time only takes 1 minutes for 100 epochs.
The model size of \netname\ is only 39 K, which is light-weighted enough with great potential for deployment on mobile devices.

\section{EVALUATION}
\label{sec:eva}

We conduct a series of experiments to evaluate the performance of \systemname. 
The dataset metrics in Section~\ref{sec:dataset}.
Since the 5G data is less, we mainly use the 4G LTE data for evaluation.
We first evaluate its performance on weather type identification, and then analyze the influence of different system design modules on the final results.

\subsection{Baseline accuracy.}
We test the ability of \systemname\ to identify rainfall utilize five-fold cross validation.
Specifically, we made two different types of splits for the dataset, including \textit{unshuffled} and \textit{shuffled}.
(1) \textit{Unshuffled}: We built the dataset in time order (the later the data point is, the larger its index in the dataset); 
(2) \textit{Shuffled}: We built the dataset in random order without considering time.
After clustering LTE data into 100, we formed a complete graph with 9 adjacent nodes for training \netname .

\begin{figure}[t]
    \centering
    \begin{minipage}{0.48\linewidth}
         \includegraphics[width=\linewidth]{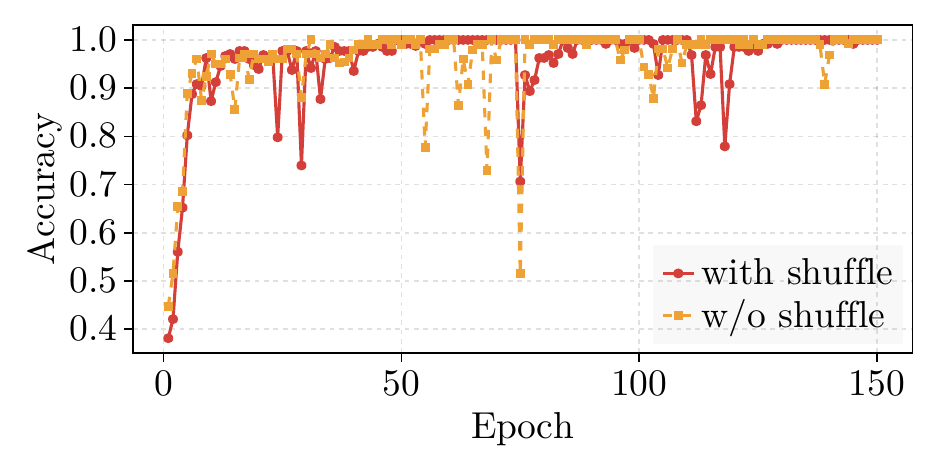}
    \caption{After approximately 200 iterations, \netname\ will converge.}
    \label{fig:time_epoic}
    \end{minipage}
    \hfill
   \begin{minipage}{0.47\linewidth}
         \includegraphics[width=\linewidth]{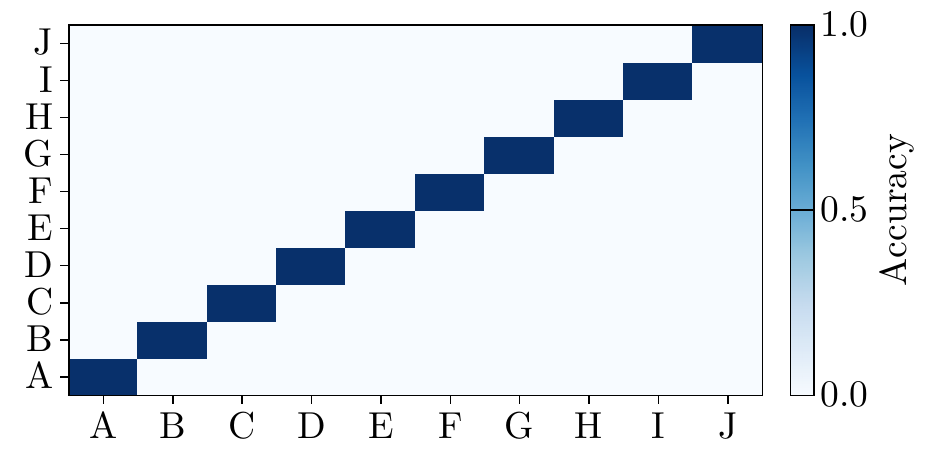}
    \caption{The accuracy rate of identification exceeds 97\%.}
    \label{fig:time_acc}
    \end{minipage}
\end{figure}

For each validation, we trained for 150 epochs, and the accuracy of the test set as a function of epochs is shown in Figure~\ref{fig:time_epoic}.
The results show that the sensing results of both shuffled and unshuffled tend to be stable after about 100 rounds of training.
There is no significant difference in the convergence speed or final accuracy between scrambled and unscrambled data during training, which we believe is due to the fact that the geographical environment around the base station remains basically unchanged within a relatively short period of time (a few hours), so that the random characteristics of the LTE signal are stable.
Figure~\ref{fig:time_acc} shows the classification results of a for ten types of rainfall data, with an average accuracy rate exceeding 98\%.
In this context, ``A" to ``J" represent a radar data interval, which is divided equally.
The maximum rainfall was moderate rain (10 $\sim $ \SI{24.9}{mm} of precipitation within 24 hours).
Compared to previous excellent works that can only identify rainfall types, such as ``light rain" and ``moderate rain", a has much finer granularity.

\subsection{The effect of the number of nodes.}
To suppress the interference caused by accidental factors such as antenna moisture, we use a complete graph composed of several nodes in the environment for rainfall sensing.
We evaluate the effect of the number of nodes in the graph on the sensing accuracy.
Specifically, the number of nodes varies from 2 to 8.
For each case, we construct datasets separately, using 50\% of them for network training and then using the remaining 50\% for testing.
The number of training epochs is 150.
Five trainings and tests are performed independently, and their average values are calculated.

\begin{figure}[t]
    \centering
    \begin{minipage}{0.48\linewidth}
        \includegraphics[width=\linewidth]{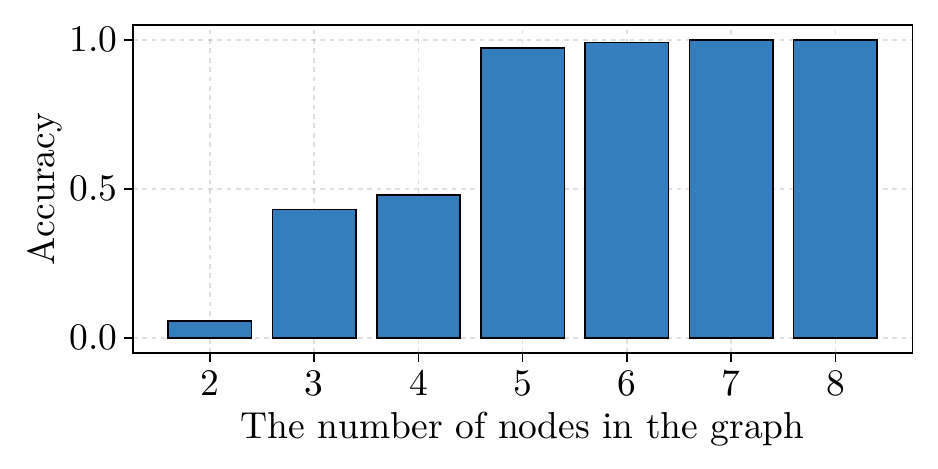}
    \caption{Increasing the number of nodes in the graph is more conducive to \netname\ extracting stable features.}
    \label{fig:time_node}
    \end{minipage}
    \hfill
    \begin{minipage}{0.48\linewidth}
        \includegraphics[width=\linewidth]{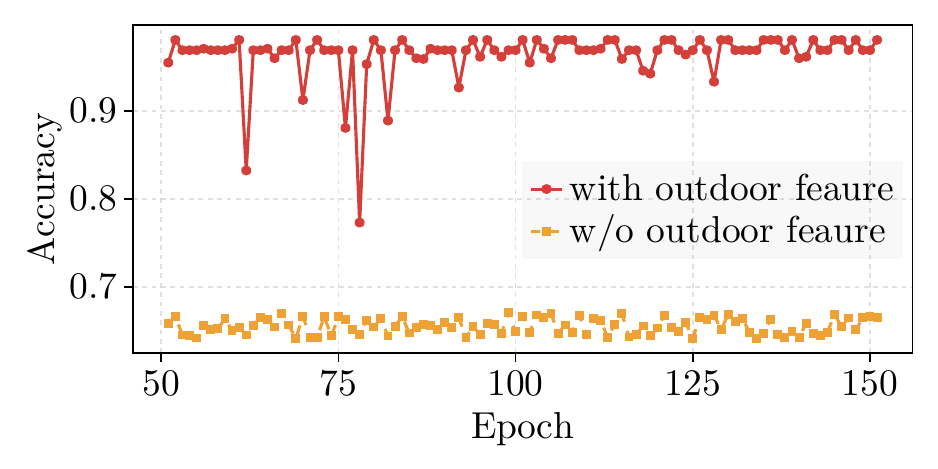}
    \caption{Constructing features related to human behavior helps achieve more accurate rainfall identification.}
    \label{fig:time_outdoor}
    \end{minipage}
\end{figure}

The results are shown in Figure~\ref{fig:time_node}.
The accuracy of identification is low (about 10\%) when only 2 nodes are used to form a complete graph.
The accuracy of identification increases with the number of nodes in a graph.
We believe this is due to the similarity of the weather in a region of several square kilometers.
According to the law of large numbers, as the number of random variable samples increases, the statistical characteristics of the sampling sequence tend to approach the actual situation of the random sequence. 
Since we use the statistical characteristics of random variables as node features for rainfall recognition in graphs, \textbf{by using multiple nodes to form a graph, \netname\ extracts overall features from it, reducing the impact of local factors (such as antenna dampness) on the identification results and improving the overall stability of the system.}

\subsection{Impact of human activity features.}
Compared with traditional rainfall monitoring methods such as rain gauges and meteorological radar, LTE data contains information related to human activities, which is related to rainfall. The messages from the base station contain information about whether the terminal equipment is indoors or outdoors, which is related to human activities.
We evaluated the impact of this information on the accuracy of rainfall identification.
Specifically, we cluster LTE data into 100 categories and then form a complete graph with adjacent 9 nodes for training.
We consider two types of node features: (1) including outdoor information, in which the node feature dimension is 16, including the proportion of outdoor terminals (POU) and the probability density of RSRP, SINR, and RSSI respectively; (2) not including outdoor information, where the diluted node feature dimension is 15, that is, only the probability density of RSRP, SINP, and RSSI without the proportion of outdoor terminals (POU).
We use 50\% of the data for training and the remaining 50\% for testing. We perform 150 rounds of training each time.

The results are shown in Figure~\ref{fig:time_node}.
The results indicate that after approximately 100 rounds of training, when the node features include POU, the accuracy exceeds 93\%, while the recognition accuracy for features without POU is less than 70\%.
We believe this is due to the strong correlation between the distribution of terminals and rainfall. \textbf{This also suggests that using communication devices such as LTE base stations to build a unified sensing system for ubiquitous sensing has advantages not possessed by traditional sensing methods.}

\subsection{Case study.}
Energy saving has received more and more attention due to global climate change.
In addition, if the base station power consumption can be dynamically adjusted by using rainfall information, it will bring huge economic benefits.
In this section, \textbf{we investigate the effect of rainfall information on energy saving.}
 Simulation results show that for a rainy region, It can solve more than \textbf{40\%} of the energy demand in the same period compared to a strategy without rainfall information.

\begin{figure}[t]
    \centering
    \begin{minipage}{0.48\linewidth}
        \includegraphics[width=\linewidth]{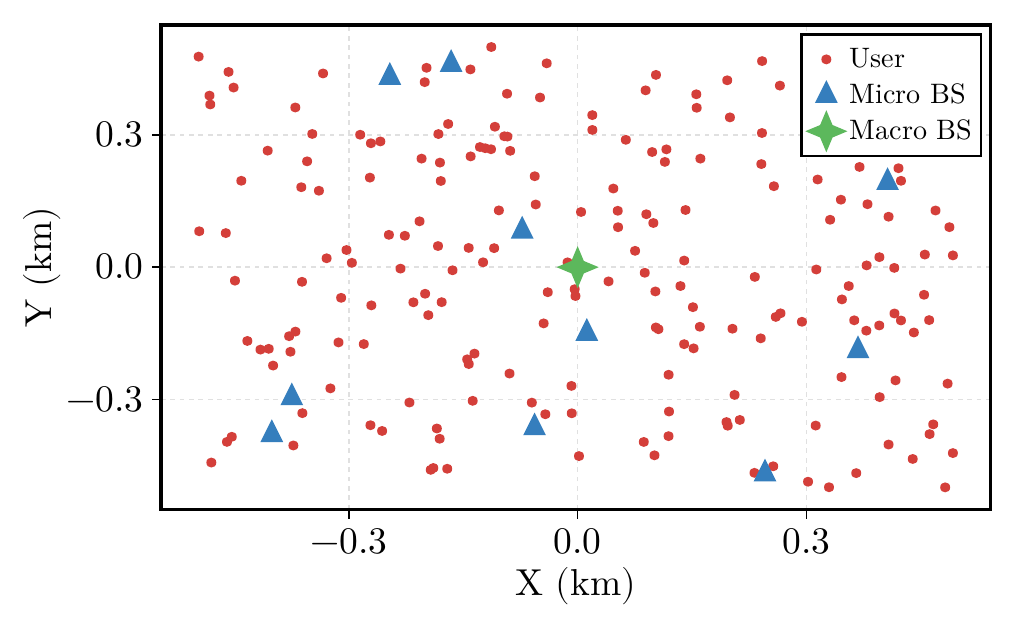}
    \caption{Distribution of base stations and users.}
    \label{fig:energy_exp}
    \end{minipage}
    \hfill
    \begin{minipage}{0.48\linewidth}
        \includegraphics[width=\linewidth]{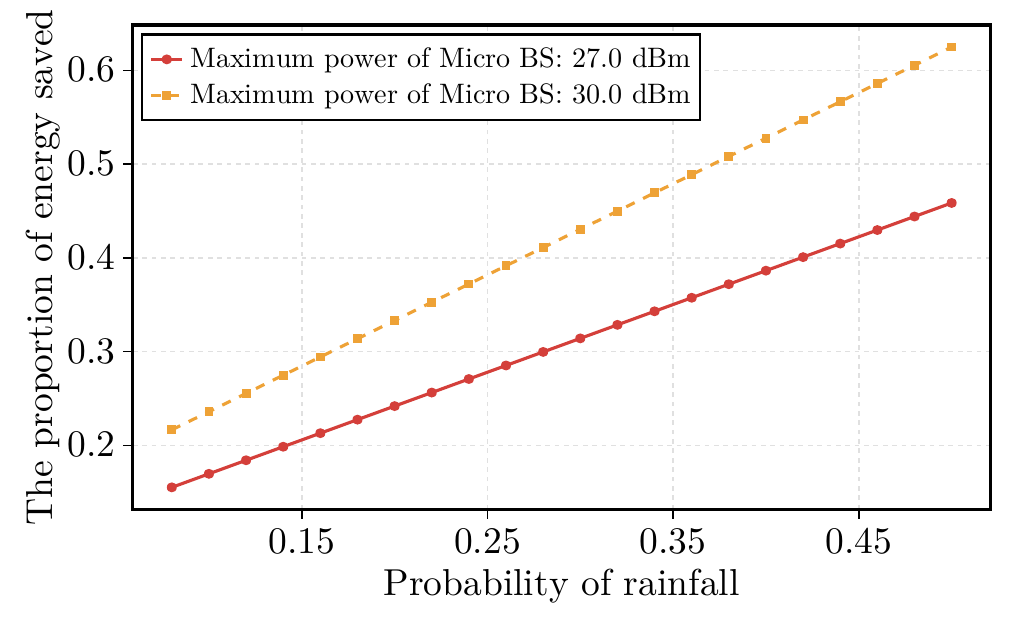}
    \caption{As the probability of rainfall increases, rainfall information will help base stations save more energy.}
    \label{fig:energy}
    \end{minipage}
\end{figure}

\textbf{Details of the simulation setup.}
In order to enhance signal coverage efficiency, current base station deployment often adopts a form of one high-power macro base station paired with several low-power micro base stations.
We consider an outdoor area as shown in Fig.~\ref{fig:energy_exp}.
A macro base station is located at the center of a \SI{1}{km} $\times$ \SI{1}{km} area, and there are 20 micro base stations uniformly distributed in the area (their locations are generated by random samplers with uniform distribution). There are 200 users uniformly distributed in the area.
We set the center frequency of the base station signal to \SI{3.4}{GHz}, which is commonly used in 5G LTE signals.
When there is no rain, the signal fading models of macro base stations and micro base stations (including path fading and shadow fading) follow the recommendations given by 3GPP 38.901 v18.0.0~\footnote{For computational convenience, we set the height of user devices to a fixed value of \SI{1.5}{m}.}.
The path loss $PL$ is given by
\begin{equation}
    PL = \Pr_{los} \cdot PL_{los} + (1-\Pr_{los}) \cdot PL_{nlos},
\end{equation}
where $\Pr_{los}$ is the probability of line-of-sight (LOS) transmission, $PL_{los}$ and $PL_{nlos}$ are the path loss of LOS and non-LOS (NLOS) transmission, respectively.
When it rains, we superimpose a rain attenuation that follows the Gaussian distribution $\mathcal{N}(9,1)$ on the base stations, where the mean value is set according to the measurement results of Beritelli et al~\cite{beritelliNeuralNetworkPattern2017}.

We set the power of the base stations to the minimum power that can meet all user communication needs (the RSSI is not less than the threshold $R$).
Specifically, if it is impossible to sense whether it is raining or not, we need to provide a configuration strategy that satisfies both rainy and non-rainy scenarios.
It is the solution $P_{wo}$ to the following optimization problem:
\begin{equation}
    \begin{aligned}
    \min_{P_{\text{ma}},P_{\text{mi}}} & \quad P_{\text{ma}}+\sum_{j=1}^{N_{mi}}P_{\text{mi},j} \\
    \text{s.t.} & \quad \text{RSSI}_{c_i,u_k}\geq R, \quad  \forall c_i \in C , u_k \in U\\
    \end{aligned}
\end{equation}
where $P_{\text{ma}}$ and $P_{\text{mi},j}$ are the power of the macro base station and the $j$-th micro base station, respectively.
$N_{mi}$ is the number of micro base stations.
$C$ is the set of weather scenarios, and $U$ is the set of users.

Fortunately, \systemname\ can realize the rain sensing at the base station, which means that we can configure different power output strategies for different scenarios to solve energy.
Specifically, in the rainfall scenario, the energy allocation strategy $P_w^{c_1}$ is the solution to the following optimization problem:
\begin{equation}
    \begin{aligned}
    \min_{P_{\text{ma}}^{c_1},P_{\text{mi}}^{c_1}} & \quad P_{\text{ma}}^{c_1}+\sum_{j=1}^{N_{mi}}P_{\text{mi},j}^{c_1} \\
    \text{s.t.} & \quad \text{RSSI}_{c_1,u_k}\geq R, \quad   u_k \in U.\\
    \end{aligned}
\end{equation}
where $c_1$ is the rain scenario.
Similarly, the configuration strategy $P_{w}^{c_2}$ in non-rainy conditions is given by the following problem, 
\begin{equation}
    \begin{aligned}
    \min_{P_{\text{ma}}^{c_2},P_{\text{mi}}^{c_2}} & \quad P_{\text{ma}}^{c_2}+\sum_{j=1}^{N_{mi}}P_{\text{mi},j}^{c_2} \\
    \text{s.t.} & \quad \text{RSSI}_{c_2,u_k}\geq R, \quad   u_k \in U.\\
    \end{aligned}
\end{equation}
where $c_2$ is the non-rain scenario.
And the total energy $P_w$ is given by
\begin{equation}
    P_w = P_{w}^{c_1} \cdot \Pr_{\text{rain}} + P_{w}^{c_2} \cdot (1-\Pr_{\text{rain}}),
\end{equation}
where $\Pr_{\text{rain}}$ is the probability of rain.

\textbf{Results.}
We set the threshold $R$ to \SI{-110}{dBm}, which is the recommended value for ensuring communication on mobile devices.
We set the maximum gain of the macro base station to \SI{53}{dBm}, which is the reference value provided by China Mobile.
The results are shown in Figure~\ref{fig:energy}.
The results show that the greater the probability of rainfall, the more energy is saved.
In particular, when the probability of rainfall is 0.5, the energy saved is more than 40\%.
We believe that the rainfall information provided by \systemname\ will provide opportunities for designing energy scheduling strategies with finer granularity and better performance.
\section{Related work}
\label{sec:related_work}

\subsection{Integrated sensing and communication}
With the development of wireless technology, more and more applications require high-quality communication and high-precision sensing at the same time~\cite{daiSurveyIntegratedSensing2022,liuIntegratedSensingCommunications2022,shangLimitsSensingCapability2024}.
The current ISAC systems primarily come in two forms. 
(1) Ubiquitous sensing using communication signals, such as WiFi, RFID, etc.
Thanks to the ubiquity of wireless signals, they exhibit superior performance in low-light and obstructed environments.
Researchers have explored their extensive applications in fields like positioning and tracking~\cite{abuhoureyahFreeDeviceLocation2023,alvarezlopezReceivedSignalStrength2017,luo3DBackscatterLocalization2019,Tamera}, material identification~\cite{liang2021fg,shang2022liqray,shangLiquImagerFinegrainedLiquid2024,wang2017tagscan,yanWiPainterFinegrainedMaterial2023}, health detection~\cite{qian2018widar2,wang2016human,yanFreeLocWirelessBasedCrossDomain2024,zhang2017toward,zhang2021widar3}, etc.
However, these systems tend to be a lack of interaction with the communication system.
(2) Sensing to assist communication.
Many excellent works~\cite{esrafilianLearningCommunicateUAVAided2019,liu2022survey,mohdnoorRemoteSensingUAV2018,nagarajanPerformanceAnalysisUAVenabled2020,ramezaniRISEnhancedIntegratedTerrestrial2022,yuanIntegratedSensingCommunicationAssisted2021} utilize communication links to estimate the target state, including position, speed, etc., thereby assisting the system in optimizing communication performance.
Such techniques are commonly applied in fields such as vehicle-road collaboration and drone platforms.
However, these systems often focus more on the impact of changes in device status (position, orientation, etc.) on the channel, while neglecting the interference caused by changes in environmental factors such as weather.
\systemname\ provides the possibility for ISAC system design based on environmental information.

\subsection{Environment knowledge map}
Traditional channel modeling often relies on empirical models, which struggle to perform fine-grained optimizations for specific environments.
For instance, the path loss provided by 3GPP is typically related only to factors such as the distance between the device and the base station, and antenna height.
However, in reality, two devices at the same distance and height may experience significant differences in signal quality due to obstructions from different buildings along their paths.
Recently, many excellent systems have attempted to utilize geographic information to construct a channel knowledge map to assist in channel modeling and optimization.
Esrafilian et al.~\cite{esrafilianLearningCommunicateUAVAided2019} try to combine with the \textbf{physical environment maps} to provide environmental information for the system.
Although such a physical environment map is readily available in some regions, it does not directly reflect the intrinsic characteristics of the channel and therefore requires additional parameters to be specified, such as the dielectric properties of the environment.
Another environment knowledge map is \textbf{radio environment map}~\cite{biEngineeringRadioMaps2019,yilmazRadioEnvironmentMap2013,zhaoNetworkSupportRadio2009} (REM), which is is a spatiotemporal database of real radio scenes, providing multi-domain environmental information such as geographical features, spectral regularities and RF emissions.
However, REM does not directly reflect its inherent channel characteristics.
Furthermore, we note that these maps often only reflect static information, such as buildings, etc.
Unfortunately, dynamic factors like rainfall can significantly affect the signal. The rainfall information provided by \systemname\ helps to improve the environmental information map.

\subsection{Weather sensing based on LTE signals}
Many previous studies~\cite{messerEnvironmentalMonitoringWireless2006,micheliRainEffect4G2021} have shown that rainfall has a significant impact on LTE signals, which brings opportunities for weather awareness.
Beritelli et al.~\cite{beritelliRainfallEstimationBased2018}  uses a classification algorithm based on the historical data of a base station to achieve a rainfall (light rain, moderate rain, heavy rain) classification accuracy rate of over 96\%.
Uijlenhoet et al.~\cite{gossetImprovingRainfallMeasurement2016,overeemRetrievalAlgorithmRainfall2016} use communication links between base stations to enable the sensing of rainfall over a large area in the Netherlands.
However, most of the communication between base stations is now replaced by underground optical cables, which forces us to seek new sensing algorithms.
Conversely, a uses the downlink for rainfall sensing. With the evolution of wireless communication technologies such as 6G, the use of wireless downlinks will become increasingly widespread.

\section{Conclusion}
\label{conclusion}

In this paper, we propose \systemname, which can use the base station downlink data for rain sensing.
\systemname\ does not require any additional hardware to be added to the existing ISAC system, and it can fully compatible with the communication mode of the existing system.
To solve the device differences existing in different base stations and terminals, including antenna gain, orientation, etc., we construct features based on the statistical characteristics of received signal strength.
In addition, we design \netname\ to extract common features in an area for rainfall sensing, which can suppress interference caused by accidental factors such as wet antennas.
Furthermore, human activity data are extracted from base station data to enhance sensing effects. We test with real data from Badaling Town, Beijing, and the results show that a can recognize ten types of rainfall with an accuracy rate of over 97\%.
We conduct a case study and the results show that the rainfall information provided by \systemname\ can save energy consumption of communication system.
We believe that it can provide new opportunities and possibilities for ISAC systems.

\bibliographystyle{acm}
\bibliography{ref,zotero}  

\end{document}